\documentclass[twoside]{dis09}
\usepackage[latin1]{inputenc}
\usepackage[dvips]{graphicx,epsfig,color}
\usepackage{wrapfig,rotating}
\usepackage{amssymb,amsmath,array}

\begin{document}
\title{Heavy flavours: working group summary}

\author{Ahmed Ali$^1$, Leonid Gladilin$^2$ and Diego Tonelli$^3$
%
%
\vspace{.3cm}\\
%
1- Deutsches Elektronen-Synchrotron DESY \\
Notkestrasse 85, D-22607 Hamburg - Germany
%
\vspace{.1cm}\\
2- Moscow State University - Scobeltsyn Institute of Nuclear Physics \\
1(2), Leninskie gory, GSP-1, Moscow 119991 - Russia
%
\vspace{.1cm}\\
3- Fermilab \\
P.O.\ Box 500, Batavia, IL 60510-5011 - USA
}

\maketitle

\begin{abstract}
The talks presented in the working group ``Heavy flavours''
of the DIS$\,$2009 workshop are summarised. New and recently
updated results from theory, proton antiproton and heavy ion colliders,
as well from HERA and $e^+e^-$ colliders are discussed~\cite{url}.
\end{abstract}

\section{Introduction}

Production and fragmentation of heavy quarks is one of the most dynamical
fields of research in QCD precision tests. There is an active interplay between
experimental analysis and theoretical developments in this area.
The aim of the working group was to discuss the relevant experimental and
theoretical results that became available or updated since the previous
DIS workshop in 2008. They are summarised here in three sections: news from 
theory (section 2), news from proton and heavy ion colliders (section 3), and
news from HERA and $e^+e^-$ colliders (section 4).

\section{News from theory}

Theoretical activity in heavy flavours has been brisk during the last
year, spurred by new experimental measurements and some outstanding
challenges. Theory talks given in this session were invited on the
 following topics:
\begin{itemize}
\item Applications of the General-Mass Variable-Flavour-Number Scheme (GM-VFNS).
\item Theoretical progress in structure functions (SFs), parton distribution
functions (PDFs), fragmentation functions (FFs) and heavy flavour production
 cross sections.
\item Studies of the $(J/\psi, \Upsilon)$ data on the production cross sections
 and polarisation.
\item Spectroscopy of the $b$-baryons and interpretations of the states
 $X(3872)$, $Y(4140)$, and $Z^+(4430)$.
\end{itemize} 
These topics were addressed in 13 theory talks.
 In this section, salient features of
the presentations  are summarised and briefly commented. 

\subsection{ Applications of GM-VFNS}
In heavy quark production
processes with two large scales, $m$, the heavy quark mass, and $p_T$, the
transverse momentum of the produced heavy quarks, one would like to absorb the
logarithms $\ln (p_T/m)$ present in the perturbative framework,
which are large in the region $p_T \gg m$, into the PDFs and FFs,
where they are resummed using the DGLAP evolution as in the massless case. 
However, one would also like to retain the finite mass terms proportional to
 $m^2/p_T^2$  in the hard-scattering cross sections.
The GM-VFNS enables to accomplish these goals in a two-step
procedure: In the first step, massless and massive quark calculations are
matched. The key point is to isolate the collinear singularities in
the massive calculation $d\tilde{\sigma}(m)$,
which are cancelled  by subtracting the massless calculation
 (in the $\overline{\rm MS}$ scheme), $d\hat{\sigma}_{\rm \overline{MS}}$, 
 but the residue $d\sigma_{\rm sub}$, defined
 as $d\sigma_{\rm sub}= \lim_{m \to 0}d\tilde{\sigma}(m)
 - d\hat{\sigma}_{\rm \overline{MS}} $, contains finite terms. The second step is
to subtract $d\sigma_{\rm sub}$ from $d\tilde{\sigma}(m )$, the cross section
 with  $m\neq 0$, in which the ultraviolet and collinear singularities due
 to massless
partons have also been removed using the $\overline{\rm MS}$ factorisation scheme.
 The resulting cross section
 $d\hat{\sigma}(m)=d\tilde{\sigma}(m) - d\sigma_{\rm sub}$ is the
desired quantity, in which finite quark mass terms are kept in the hard
 scattering
cross section but the PDFs and FFs for the massless partons in the
 $\overline{\rm MS}$ scheme are used. Details can be seen in~\cite{Kniehl:2005mk}.
 The  GM-VFNS approach has been applied to a
number of inclusive one-particle production cross sections in the processes
 $\gamma + \gamma \to D^{*\pm} +X $,  $\gamma^* + p \to D^{*\pm} +X $, 
$p + \bar{p} \to (D^0, D^{*\pm}, D^\pm, D_s^\pm, \Lambda_c^\pm) + X $
and $p + \bar{p} \to (B^0, B^\pm) +X$. Of these, the photoproduction
  $\gamma^* + p \to D^{*\pm} +X $, open charm hadroproduction
$p \bar{p} \to D+X$, and heavy quark electroproduction are being discussed in
 detail in these proceedings by 
Spiesberger~\cite{spiesberger}, Kniehl~\cite{kniehl}, and
 Alekhin~\cite{alekhin},   respectively. In photoproduction, one has to treat
 consistently the contributions from 
the direct and resolved processes. At the next-to-leading order (NLO)
 theoretical
precision, the separation into direct and resolved contributions becomes
scheme-dependent, as the singular contributions to the direct photon
part have to be factorised and absorbed into the parton distribution
functions of the photon. Data on
photo- and electro-production of $D^*$ meson from H1 and ZEUS 
have been analysed in the GM-VFNS framework~\cite{spiesberger}. There is good 
overall agreement between data and theory, but the theoretical uncertainties 
are still large, dominated by the ambiguity in the choice of the
renormalisation and factorisation scales. In the small-$p_T$ region, 
theoretical prediction is uncertain by typically a  factor 2  which then is
propagated to the entire
rapidity distributions, as they were measured with a relatively small lower
cuts, $p_T> 1.8$ GeV for the photoproduction data from H1 and $p_T > 1.5$ GeV
for low-$Q^2$ data from ZEUS. A NNLO calculation is required to reduce these
uncertainties. A Monte Carlo programme for heavy quark production based on
the NLO calculations has also been developed and is now in use in the
analysis of the HERA data~\cite{toll}.

 Transverse-momentum distributions for the charmed hadrons in
$p \bar{p} \to (D^0,D^+,D^{*+}) + X$  are discussed in
\cite{kniehl}. The essential improvement here is a better determination of the
FFs using the LEP1, KEKB and CESR data in the GM-VFNS scheme, which has 
resulted in better agreement of the $p_T$-distributions of the charmed
hadrons at the Tevatron. In the GM-VFNS scheme, D-meson hadroproduction 
receives contributions from the partonic processes with incoming $c$ and
$\bar{c}$, and hence this process is sensitive to their PDFs and possible
intrinsic charm (IC)-induced enhancements. The IC contribution has been
studied in a variety of different models and analysed in~\cite{Pumplin:2007wg}
using the CTEQ6.5 global analysis~\cite{Tung:2006tb}. The $p_T$-distributions
in $p\bar{p} \to D^0 +X$ at the Tevatron and in $p p \to D^0 +X$ at RHIC
(with $\sqrt{s}=200$ GeV and $\sqrt{s}=500$ GeV) have been calculated in the  
GM-VFNS scheme and the former are compared with the Tevatron data, using six
 different
parameterisations of the IC-contribution. Unfortunately, no firm conclusions
can be drawn as the present Tevatron data, which is of the 2002
vintage, is based on rather modest luminosity (5.8 pb$^{-1}$). This will
change greatly if the full force of the current Tevatron integrated
luminosity (circa $5\,$fb$^{-1}$) is brought to bear on this analysis.
Data at RHIC are also
potentially sensitive to the input IC-contribution and will be able to
discriminate among the various IC-models in the future. 

The charm structure function $F_2^{c{\bar c}}(x,Q^2)$ measured in the DIS
electroproduction at HERA was discussed by Alekhin~\cite{alekhin} in the
 context of the two
schemes FFNS (fixed flavour-number scheme) and the VFNS, using the
Buza-Matiounine-Smith-van Neerven (BMSN) prescription~\cite{Buza:1996wv}.
The main conclusion from this analysis is that for the realistic HERA
 kinematics,
 the two schemes are rather similar. FFNS with partial $O(\alpha_s^3)$
corrections incorporated provides a good description of the HERA data for
small/moderate $Q^2$ values, but it undershoots the data for high $Q^2$,
calling for including the remaining $O(\alpha_s^3)$ pieces. Finally, an
intermediate mass scheme was discussed by Nadolsky~\cite{nadolsky} and a
global PDF analysis was carried out in this scheme with the conclusion 
that the intermediate-mass formulation improves the NLO zero-mass scheme
and approximates the more fundamental General-mass scheme
(GM-VFNS) in a simple way.   

\subsection{Progress in structure functions involving heavy quarks}
State of the art QCD analysis requires the description of the heavy quark
contribution to the structure functions at 3 loops to match the accuracy
reached in the massless partonic case. Making use of the factorisation
property of the heavy quark Wilson coefficients, denoted by $H_{(2,L),j}$,
into a product of the massive operator matrix elements $ A_{kj}^{S, NS}$, and
 the light flavour Wilson coefficients $C_{(2,L),k}^{S, NS}$, which holds in
 the region $Q^2 \gg m^2$, one has
\begin{equation}
H_{(2,L),j}^{S,NS}\left(\frac{Q^2}{\mu^2}, \frac{m^2}{\mu^2}\right) = 
A_{kj}^{S,NS}\left(\frac{m^2}{\mu^2}\right) \otimes C_{(2,L),k}^{S,NS}
\left(\frac{Q^2}{\mu^2}\right)~.
\end{equation}
Analytic results are known for $Q^2 \gg m^2$ at NLO for $F_2(x,Q^2)$ and
at NNLO for $F_L(x,Q^2) $. The light flavour  Wilson coefficients are also
known at the NNLO, thanks to the formidable calculation by Vermaseren, Vogt and
Moch~\cite{Vermaseren:2005qc}. At this conference, massive operator matrix elements 
$A_{kj}^{S,NS}(\frac{m^2}{\mu^2})$, contributing to $F_2(x,Q^2)$ were reported
in the region $Q^2/m^2 \geq 10$ to 3-loop accuracy for the fixed moments
of the Mellin variable $N$~\cite{klein}. In a computational
 {\it Tour de force}, the following fixed moments of the massive operator
matrix elements were accomplished~\cite{Bierenbaum:2009mv}:
\begin{eqnarray}
&A_{Qq}^{(3), {\rm PS}}:& (2,4,...,12);~~ A_{qq,Q}^{(3), {\rm PS}}, A_{gq,Q}^{(3)}:
 (2,4,...,14)~;
\nonumber\\
&A_{qq,Q}^{(3),{\rm NS}\pm}:& (2,3,...,14); ~~ A_{Q(q)g}^{(3)}, A_{gg,Q}^{(3)}:
 (2,4,...,10)~.
\end{eqnarray} 
The superscript ${\rm PS (NS)}$ stands for the pure singlet (non-singlet) case, and the
quarkonic operator matrix element is represented as
 $A_{qq}^S= A_{qq}^{NS}+ A_{qq}^{PS}$. In the case of the flavour non-singlet 
contributions, also the odd moments are calculated. This provides an
independent check on the 3-loop anomalous dimensions $\gamma_{qg}^{(2)}$,
$\gamma_{qq}^{(2),PS}$ and on the respective colour projections of
$\gamma_{qq}^{(2),NS\pm}$,   $\gamma_{gg}^{(2)}$ and $\gamma_{gq}^{(2)}$.
Phenomenological parameterisations are to follow soon.

Heavy flavour effects in the virtual photon structure functions in the NLO
accuracy were reported at this meeting by Uematsu~\cite{uematsu}. Concentrating
on $F_2^\gamma(x,Q^2,P^2)$, where $Q^2$ is the photon virtuality and
$P^2$ the target (photon) mass, one has the factorisation result as
a convolution in the virtual photon distribution function $\vec{q}^\gamma$
and the coefficient function $\vec{C}$ 
\begin{equation}
F_2^\gamma(x,Q^2,P^2) = \vec{q}^{\gamma}(y,Q^2,P^2,m^2) \otimes
 \vec{C} \left(\frac{x}{y},\frac{\bar{m}^2}{Q^2},\bar{g}(Q^2)\right)~, 
\end{equation}
where the heavy quark mass dependence enters in both the terms. Taking the
 moments of  $\vec{q}^\gamma$, one can write the resulting expression as a
product of two terms
\begin{equation}
\int_0^1 dx x^{n-1} \vec{q}^\gamma (x,Q^2,P^2,m^2) =
\vec{A}_n\left(1,\frac{\bar{m}^2(P^2)}{P^2}, \bar{g}(P^2)\right) T \exp \left[
\int_{\bar{g}(Q^2)}^{\bar{g}(P^2)} dg \frac{\gamma_n(g,\alpha)}{\beta(g)}
 \right],
\end{equation}
where the matrix element
\begin{equation}
\langle \gamma(P^2)|\vec{O}_n (\mu^2)|\gamma(P^2) \rangle
=\vec{A}_n \left(\frac{P^2}{\mu^2}, \frac{\bar{m}^2(\mu^2)}{\mu^2}, \bar{g}
 (\mu^2)\right)
\end{equation}
is perturbatively calculable. Thus, as opposed to the nucleon structure
 functions discussed above, the photon structure function is completely
calculable perturbatively, as first emphasised by Witten~\cite{Witten:1977ju}.
This framework has been applied in the massive quark limit
$\Lambda_{\rm QCD}^2 \ll P^2 \ll m^2 \ll Q^2$ to the PLUTO data with
$Q^2=5$ GeV$^2$ and $P^2=0.35$ GeV$^2$, with the conclusion that the data 
is in better agreement
with the case of 3 massless NLO QCD $+$ $c$ massive compared to the
4 massless NLO QCD case~\cite{Kitadono:2008iw}. A comparison of the massive
 $b$ quark + 4 massless
quarks case with the L3 data for $Q^2=120$ GeV$^2$ and $P^2=3.7\,$GeV$^2$ is
not as conclusive due to the smaller $(-1/3)$ electric charge of the $b$-quark
as well as the imprecise nature of the L3 data. 

\subsection{Progress in the $t\bar{t}$ production cross section at the Tevatron
 and the LHC}
There has been steady progress in calculating the $t\bar{t}$ production cross
 section at the Tevatron and the LHC. This was reported at this conference by
Langenfeld~\cite{langenfeld}, and since then in a recently published
 paper~\cite{Langenfeld:2009wd}, in which the cross section at the Tevatron is
used to determine the running top quark mass. The relevant formulae for 
the process $\sigma (p p (\bar{p}) \to t \bar{t} +X)$ are:
\begin{equation}
\sigma (pp(\bar{p}) \to t\bar{t}+X)= \frac{\alpha_s^2}{m_t^2} 
\sum_{i,j=q, \bar{q}, g} \int_{4m_t^2}^S ds L_{ij}(s,S,\mu_f^2) f_{ij}(\rho,M,R)~,
\end{equation}
\begin{equation}
L_{ij}(s,S,\mu_f^2)= \frac{1}{S} \int_s^S \frac{d\hat{s}}{\hat{s}}
 \phi_{i/p}\left(\frac{\hat{s}}{S},\mu_f^2\right)
\phi_{j/p(\bar{p})}\left(\frac{s}{\hat{s}},\mu_f^2\right)~,
\end{equation}
Here, $S$ is the centre-of-mass energy squared (of the $pp$ or $p\bar{p}$
colliding beams), $s$ is the partonic centre-of-mass energy squared, $m_t$ is
the top quark (pole) mass, and $L_{ij}$ is the parton
luminosity function with the PDFs $\phi_i/p$, evaluated at the factorisation
scale $\mu_f$.  The functions $f_{ij} (\rho,M,R)$ are the hard partonic
cross section and depend on the dimensionless variables $\rho =4m_t^2/s$,
$R=\mu_r^2/\mu_f^2$ and $M=\mu_f^2/m_t^2$, where $\mu_r$ is the renormalisation scale.
They have been solved in perturbation theory up to two loops. For $M=R=1$,
i.e., $m_t=\mu_r=\mu_f$, one has $f_{ij}(\rho,1,1)=f_{ij}^{(0)}(\rho) +
 4\pi \alpha_s f_{ij}^{(1)}(\rho) + (4\pi \alpha_s)^2 f_{ij}^{(2)}$. The 
dependence of $f_{ij} (\rho,M,R)$ on the scales $\mu_f$ and $\mu_r$ can be
made explicit and the expressions can be seen in~\cite{Langenfeld:2009wd}.
Up to one loop, the results are known since long, and at the two-loop level,
the corrections include the Sudakov logarithms $\ln^k\beta$, where $\beta$ is
the $t$-quark velocity, $\beta=\sqrt{1-\rho}$, and terms with $k=1,...,4$ are
included in $f_{q\bar{q}}^{(2)}$ and $f_{gg}^{(2)}$ (the leading term 
$\sim \ln^3\beta$ in $f_{qg}^{(2)}$), as well as the Coulomb contributions
 $\sim1/\beta^2, 1/\beta$. The subleading correction in the function
$f_{ij}^{(2)}$ has yet to be calculated. Estimating it as $O(30\%)$ in the NNLO
contribution implies an uncertainty of $\Delta \sigma(pp \to t\bar{t}+X)=
{\cal O}(15)$ pb at the LHC and $\Delta \sigma(p\bar{p} \to t\bar{t}+X)=
{\cal O}(0.2)$ pb at the Tevatron. The rest of the uncertainties come from the
PDFs, the scales $\mu_f$ and $\mu_r$, $\alpha_s$ and $m_t$. For the pole
mass $m_t=173$ GeV,
the $t\bar{t}$ cross sections at the LHC ($\sqrt{S}=14$ TeV) and the
 Tevatron ($\sqrt{S}=1.96$ TeV) are estimated (for the MSTW2008 PDFs~\cite{Martin:2009iq})
 as follows~\cite{Langenfeld:2009wd}:
\begin{equation}
\sigma (pp \to t\bar{t} +X)_{\rm LHC}= (887 ^{+9}_{-33} ({\rm scale}) \pm 15
({\rm PDF}))~{\rm pb}~,
\end{equation}
\begin{equation}
\sigma (p\bar{p} \to t\bar{t} +X)_{\rm Tevatron}= (7.04^{+0.24}_{-0.36}
 ({\rm scale}) \pm 0.14 ({\rm PDF}))~{\rm pb}~.
\end{equation}
For the CTEQ6.6 set of PDFs~\cite{Nadolsky:2008zw}, the central values of the cross section 
become 874 pb and 7.34 pb, for the LHC and the Tevatron, respectively,
with almost the same scale uncertainties as for the MSTW2008 PDFs, but
the PDF-related errors for this set are significantly larger, $\pm 28$ pb in
$\sigma_{\rm LHC}$ and $\pm 0.41$ pb in $\sigma_{\rm Tevatron}$, respectively.
Finally, using the well-known relation between the pole mass $m_t$ and the
$\overline{\rm MS}$ mass $\bar{m}_t(\mu_r=\bar{m}_t)$, and making the $m_t$-dependence in the
total cross section manifest, a value $\bar{m}_t(\bar{m}_t)=160.0^{+3.3}_{-3.2}$ GeV for the
 $\overline{\rm MS}$ top quark mass is obtained from the Tevatron
 production cross section.

\subsection{Theoretical status of $J/\psi$ production at HERA}
The cross section $\sigma(\gamma p \to J/\psi +X)$ in the colour-singlet
(CS) model at the NLO accuracy was calculated some time ago by 
Kr\"amer~\cite{Kramer:1995nb} and has been reconfirmed recently in 
\cite{Artoisenet:2009xh}, in which polarisation observables 
in photoproduction were also calculated at the NLO accuracy. These
calculations have been compared with the published~\cite{Chekanov:2002at}
 and preliminary ZEUS data~\cite{brugnera:2008} and the results are summarised at
this conference by Artoisenet~\cite{artoisenet}. The 
hadronic matrix element $\langle {\cal O}_{J/\psi}(^{3}S_1)[1]\rangle$
 was fixed from the analysis of the 
$J/\psi$-hadroproduction data and the CTEQ6M PDF set~\cite{Pumplin:2002vw} was used. The three
main phenomenological parameters $m_c$, the charm quark mass, and
$\mu_r, \mu_f$, the renormalisation and factorisation scales, were varied
 in the range $1.4~{\rm GeV} < m_c <
1.6~{\rm GeV}$, $\mu_0=4 m_c$, $0.5\mu_0 < \mu_r, \mu_f < 2 \mu_0 $, and
$0.5 < \mu_r/\mu_f <2  $. It was found that the CS yield at NLO accuracy
 underestimates the ZEUS data in both the $p_T$ and $z$-distributions, where
$z=p_\psi .p_p/p_\gamma.p_p$ -- a conclusion which is at variance
with the observations made earlier~\cite{Chekanov:2002at}. The polarisation
of $J/\psi$ is studied by analysing the angular distributions of
the leptons originating from the decay $J/\psi \to \ell^+\ell^-$. In terms
of the polar and azimuthal angles $\theta$ and $\phi$ in the $J/\psi$ rest
frame, one has
\begin{equation}
\frac{d \sigma}{d \Omega dy} \propto 1 + \lambda(y) \cos^2\theta + \mu(y)
\sin 2 \theta \cos \phi + \frac{\nu(y)}{2} \sin^2 \theta \cos 2 \phi~,
\end{equation}  
where $y$ stands for either $p_T$ or $z$. If the polar axis coincides with the
spin quantisation axis, the quantities $\lambda(y)$, $\mu(y)$ and $\nu(y)$
can be related to the spin density matrix elements. The $O(\alpha_s)$
corrections in the CS model have a strong impact on the polarisation parameters
$\nu$ and $\lambda$, which were analysed as functions of $p_T$ and $z$.
In particular,  the $\lambda$-distribution in NLO decreases rapidly with
increasing $p_T$ and has a large negative value above $p_T=4$ GeV, in
contrast to the LO prediction, which is in reasonably good agreement with
the data. The measurements of the parameter $\nu$ are, on the other hand,
in better agreement with the NLO predictions (compared to the LO CS model
prediction), though the model uncertainties are too large for low $z$ values
to draw a quantitative conclusion. 

 On a related issue, Faccioli~\cite{faccioli} summarised the experimental 
situation of $J/\psi$ polarisation from the fixed target (E866, HERA-B) to the
 collider energies (CDF), observing that the magnitude and the ``sign'' of
 the measured $J/\psi$ polarisation crucially depends on the reference 
frame~\cite{Faccioli:2008dx}. In particular, he showed
that the seemingly contradictory data on the parameter $\lambda$
from the experiments E866, HERA-B and CDF overlap as a function of the
$J/\psi$ cms total momentum and the data can be consistently described, 
assuming that the most suitable  axis for the measurement is along the
 direction of the relative motion of the colliding  partons. In this
 (Collins-Soper) frame, polarisation changes from the
longitudinal at small momentum to transverse at high momentum. This resolves
the apparent $J/\psi$ polarisation puzzle among the experiments. However,
the puzzle involving the theory (based on NRQCD) vs. experiment still persists. 
 In conclusion, a quantitative understanding 
of the $J/\psi$-photoproduction data at HERA is still lacking, calling for
improved calculations within the NRQCD (such as invoking colour-octet
transitions and $O(\alpha_s^2)$ corrections to the CS model) or perhaps
the $J/\psi$ data are inviting a better theoretical framework. Soft Collinear
Effective Theory is a case in point.
A theoretical analysis of
the production cross section and the spin alignment parameter
 $\alpha(\Upsilon)$ in $p \bar{p} \to \Upsilon +X$ data from the D0
 collaboration at the Tevatron was presented by Zotov~\cite{zotov}, with
very similar conclusion, namely that the existing QCD framework is not
in agreement with the data, in particular the distribution of
 $\alpha(\Upsilon)$ as a function of $p_T$ is not understood. This will come under 
sharp scrutiny at the Tevatron and certainly at the LHC.   

\subsection{New developments in the spectroscopy of charm and beauty hadrons}
The saga of the successful predictions of the constituent quark model (CQM)
 continues!
Salient features of the $b$-baryon spectroscopy in this model were discussed
and contrasted with the existing data by Karliner~\cite{theory-omegab}, with the
conclusion that CQM (with colour hyperfine interaction) gives a highly accurate
predictions for the heavy baryon masses. Four examples from the $b$-baryon
spectroscopy illustrate this: The measured mass difference by the CDF
collaboration $m({\Sigma_b}) - m({\Lambda_b})= 192 $ MeV was predicted to be
194 MeV; the hyperfine splitting $m(\Sigma_b^*) - m(\Sigma_b) =21$ MeV [CDF]
is in agreement with the predicted value of 22 MeV. Likewise, the 
predictions for the masses, $m(\Xi_b)=5795 \pm 5$ MeV
vs.~$5793\pm 2.4 \pm 1.7$ MeV (expt.), and  $m(\Omega_b)=6052.9 \pm 3.7$ MeV,
vs.~$m(\Omega_b)=6054.4 \pm 6.8~({\rm stat})~\pm 0.9~({\rm syst.)}$ MeV [CDF],
are in excellent agreement with data. It should, however, be noted that the
measurement of the $\Omega_b$ mass by CDF~\cite{cdf-omegab} differs from the first reported
measurement of the $\Omega_b^-$ mass by D0,  $m(\Omega_b^-)= 6165 \pm 10
(\rm{stat}) \pm 13 ({\rm syst})$ MeV~\cite{Abazov:2008qm}. 
 On the theoretical side, the aspect that 
the constituent quark masses in this model, in particular the mass difference $m_b-m_c$
used as input, depend on the spectator quarks, deserves further study. 
 Thus, the relation of the CQM with QCD, in which quark masses are universal, is far
 from obvious.

The observation of the narrow state $X(3872)$ by BELLE, in the decay mode
$B^\pm \to K^\pm \pi^+\pi^- J/\psi$, with the $\pi^+\pi^-J/\psi$-mass
spectrum peaking at 3872 MeV, dominated by the state
 $X(3872) \to J/\psi \rho$, as well as the radiative decay mode
 $X(3872) \to J/\psi \gamma$, measured by  BABAR, has established 
 $X(3872)$ as an exotic $J^{\rm PC}=1^{++}$ state.
CDF and D0 confirmed the $X(3872)$ in $p\bar{p}$ collisions (produced 
predominantly in prompt processes rather than in $B$ decays). In addition,
the decay mode $X(3872) \to J/\psi \omega$, which seems to have a similar
branching ratio as the mode  $X(3872) \to J/\psi \rho$, implies large
isospin violation, not typical of the usual strong interactions.
 The two leading hypotheses are that $X(3872)$ is a loosely bound 
``hadronic molecule'' of the $D^0 \bar{D}^{0*}$, since the mass of  
$X(3872)$ is so close to the $D^0 \bar{D}^{0*}$ threshold~\cite{Thomas:2008ja},
or that it is a point-like hadron, a tightly bound state of a diquark and
an antidiquark, a tetraquark~\cite{Maiani:2004vq}.
The drawback of the tetraquark picture is that it predicts a very rich
 spectroscopy. 
However, one has little understanding why a large number of these
states have not been found. One of these predictions, namely the existence of related
charged particles decaying into charmonium and pions, is still being debated.
 The molecular picture seems 
to be at odds with the large prompt production in hadronic collisions.
 At this conference, Polosa~\cite{polosa} presented a calculation 
for the production cross section of $X(3872)$ at the Tevatron, assuming that
$X(3872)$ is a loosely bound $D^0 \bar{D}^{0*}$ molecule. Since then, this
work has been published~\cite{Bignamini:2009sk} and the main points are
summarised here. Using the CDF data on prompt $X(3872)$ production in the
$J/\psi \pi^+\pi^-$ mode and the yield of $\psi(2S) \to J/\psi \pi^+\pi^-$,
as well as the fraction  of the prompt $\psi(2S)$ candidates, and assuming that
the $X(3872)$ and $\psi(2S)$ have the same rapidity distribution in the
range $|y| <1$, a lower bound on the prompt production  cross section 
$\sigma(p\bar{p} \to X(3872) + X)$ of
$3.1 \pm 0.7$ nb is obtained. This is contrasted with the estimates of an
upper bound on the theoretical cross section, based on
the assumption that $X(3872)$ is an S-wave bound state, $X(3872) =1/\sqrt{2}
(D^0\bar{D}^{*0} + \bar{D}^0 D^{*0})$:
\begin{equation}
\sigma(p\bar{p} \to X(3872) + X) \propto |\int d^3{\mathbf k} \langle X|
D\bar{D}^{*}({\bf k})\rangle \langle D\bar{D}^{*}({\bf k})|p\bar{p}\rangle|^2
\leq \int_{{\cal R}} d^3{\mathbf k} |\langle D\bar{D}^{*}({\bf k})|p\bar{p}\rangle|^2
~,
\end{equation}  
where ${\mathbf k}$ is the relative 3-momentum between the $D({\mathbf p}_1)$
and $D^*({\mathbf p}_2)$ mesons.
 $\psi({\mathbf k})=\langle X| D\bar{D}^{*}({\bf k})\rangle$
is a normalised wave function for the state $X(3872)$, and ${\cal R}$ is
the region where $\psi({\mathbf k})$ is non-zero. The matrix element
 $\langle D\bar{D}^{*}({\bf k})|p\bar{p}\rangle$ is calculated with the
help of QCD $(2 \to 2)$ matrix elements embedded in the fragmentation programme,
Pythia and Herwig. With the binding energy
 ${\cal E}_0 \sim M_X-M_D-M_{D^*}= -0.25
\pm 0.40$ MeV, the characteristic size $r_0$ of the molecule is estimated as
$r_0 \sim (8.6 \pm 1.1)$ fm, yielding for $k$ a number
 $k \simeq \sqrt{\mu (-0.25 \pm 0.40)} \simeq 17$ MeV, where
 $\mu= m_{D}m_{D^*}/(m_D+ m_{D^*})$ is the reduced mass. Applying the uncertainty
principle yields a Gaussian momentum spread $\Delta p \sim 12$ MeV.
Alternatively, $k$ is of the order of the centre of mass momentum,
$k \simeq 27$ MeV. With this,
the integration region  is restricted to a ball ${\cal R}$ of radius 
$\simeq [0,35]$ MeV. Herwig then yields an upper limit of
$0.013$ nb on $\sigma(p\bar{p} \to X(3872) + X)$ and the corresponding number for Pythia
 is $0.036$ nb. This is 
typically two orders of magnitude smaller than the measured cross section by
 CDF, disfavouring the molecular interpretation of $X(3872)$. 

\section{News from proton and heavy ion colliders}

\subsection{New results from $p\bar{p}$ and $pp$ colliders}

The Tevatron is today the only environment where all species of heavy
flavours (HF) are studied.  The CDF and D0 experiments pursue a rich and
diverse HF program that is reaching maturity, owing to samples of
$p\bar{p}$ collisions in excess of 5 fb$^{-1}$ per experiment,  
expected to
double by 2011.
This suggests a few years of intense and fruitful competition with next
generation experiments that will soon start their operations at the LHC.

\begin{wrapfigure}{r}{0.5\columnwidth}
\centerline{\includegraphics[width=0.45\columnwidth] 
{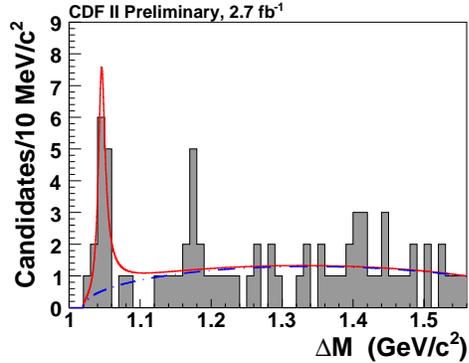}}
\caption{Difference between the $J/\psi\phi$  and the $J/\psi$ mass
as observed by CDF in $2.7\,$fb$^{-1}$ of Tevatron $p\bar{p}$ data  
with fit projection in the background-only (dashed line) and signal 
+background hypothesis (full line) overlaid.}\label{Fig:Y}
\end{wrapfigure}
With thousands of top-quark decays collected, Tevatron top physics  
entered the realm of precision~\cite{top}. Using 2.8 fb$^{-1}$ CDF  
reported the single most precise $t\bar{t}$ production cross section  
measurement, $\sigma_{t\bar{t}} = 6.9 \pm 0.4 (\mbox{stat}) \pm 0.4  
(\mbox{syst}) \pm 0.1 (\mbox{theo})$ pb; the 8\% relative uncertainty  
challenges the precision of most recent theory predictions, and is  
achieved
by normalising the $t\bar{t}$ to the $Z$ cross section thus cancelling  
the leading systematic uncertainty from the luminosity of the sample.  
D0 reports a result, combined through 14 independent channels, of $ 
\sigma_{t\bar{t}} = 8.18 ^{+0.98}_{-0.87}$ pb using 0.9 fb$^{-1}$.  
While measurements of $t\bar{t}$ strong production are an important  
test of perturbative QCD, electroweak (single) top-quark production  
determines directly the magnitude of the $V_{tb}$ quark-mixing matrix  
element, probes the $b$--quark PDF of the proton, is sensitive to  
fourth generation models, and is ultimately a key testing ground for  
Higgs searches in the \emph{WH} associated production channel. Both  
experiments reported $5\sigma$ observation of this process using 2.3  
(D0) and 3.2 (CDF) fb$^{-1}$. Because of the tiny signal-to-background  
ratio, use of advanced machine-learning techniques is required, whose  
performance is carefully validated in multiple control samples. The  
measured cross sections,  $\sigma_{t} = 2.3 ^{+ 0.6}_{- 0.5}$ pb (CDF)  
and $\sigma_{t} = 3.94 \pm 0.88$ pb (D0), yield the values $|V_{tb}| =  
0.91 \pm 0.13$ (CDF) and $|V_{tb}|=1.07 \pm 0.12$ (D0).

\begin{wrapfigure}{r}{0.5\columnwidth}
\centerline{\includegraphics[width=0.30\columnwidth] 
{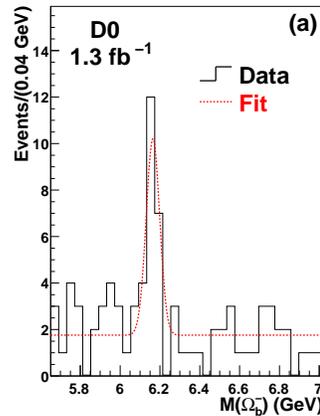}}
\caption{Mass distribution of $\Omega_b^+$ candidates reconstructed by  
D0 in 1.3 fb$^{-1}$ of Tevatron $p\bar{p}$ data with fit projection  
overlaid.}\label{Fig:Omega}
\end{wrapfigure}
Interest in HF spectroscopy was renewed recently, after a few  
unexpected ``exotic'' states, observed at the $B$ factories,  
challenged our understanding of hadrons' composition~ 
\cite{spectroscopy}. The latest addition to this picture comes from  
CDF, which reports evidence of a new resonant $J/\psi\phi$ state,   
reconstructed from the world's largest sample ($\approx 75$ events) of  
$B^+ \to J/\psi\phi K^+$ decays, in $2.7\,$fb$^{-1}$ of data. The  
excess amounts to $14 \pm 5$ events, for a significance in excess of  
$3.8\sigma$ (Fig.~\ref{Fig:Y}). The particle is dubbed $Y(4140)$ after  
its observed mass of $4143 \pm 2.9 (\mbox{stat})\pm 1.2$ (\mbox{syst})  
MeV, which is above the charm-pair threshold and disfavours  
interpretations as a conventional charmonium state. D0 reports the  
first observation of the $\Omega_b^+$ baryon (quark content $ssb$) in  
its $J/\psi(\to \mu\mu)\Omega^+[\to \Lambda(\to p\pi) K^+]$ decay with  
$18 \pm 5$ events in 1.3 fb$^{-1}$ and a significance greater than  
$5\sigma$ (Fig.~\ref{Fig:Omega}). The observed mass value,  
$m(\Omega_{b}^{+}) = 6165 \pm 10 (\mbox{stat})\pm 13$ (\mbox{syst})  
MeV, is higher than most theoretical predictions. Shortly after this  
workshop, CDF has reported observation of this baryon with mass  
$6054.4 \pm 6.8 (\mbox{stat}) \pm 0.9$ (\mbox{syst})~\cite{cdf-omegab},
in good agreement with theoretical expectations~\cite{theory-omegab}.

Hadroproduction of HF is a crucial test of our understanding of
QCD~\cite{production-1}. CDF presented a measurement of $B_c^+$
production cross section in the semileptonic $J/\psi(\to \mu^+\mu^-) 
\mu^+ X$
final state using $B^+\to J/\psi K^+$ as a reference.
The ratio of production rates times branching ratios in the $p_T>4\,$GeV
regime is $R = 0.295 \pm 0.063$, using $1\,$fb$^{-1}$. CDF also  
reconstructed a few hundred events in the exclusive channel $B^+_c \to  
J/\psi\pi^+$ using the full 4.7 fb$^{-1}$ sample. A signal with  
similar yield and purity is expected from  LHCb with just 1 fb$^{-1}$  
of data.  However, the general common strategy for initial  
measurements at LHC is to focus on either inclusive final states or  
well-known exclusive ones, like $B^+\to J/\psi K^+$. The LHC  
experiments plan to provide a deeper insight into some theory/data  
discrepancies observed in prompt \emph{onia} production (spectra and  
polarisations) at the Tevatron.  Even with a small fraction of their  
initial data, significant samples are expected, by exploiting larger  
and complementary detector acceptances with respect to CDF and D0:  
e.g. CMS expects $\approx 75,000$ $J/\psi$ decays in only 3 pb$^{-1}$  
of data.

CDF also reported the first observation  of exclusive central charmonium
~(photo)pro-duction~\cite{production-2} in hadron collisions.
Clear $J/\psi$ and $\psi(2S)$ signals over a negligible continuum  
background are reconstructed, in agreement with theory predictions.  
Observation of exclusive $\chi_{c0}$ production as well provides  
useful constraints on the reach in exclusive Higgs production at the  
LHC.

\subsection{New results from heavy ion colliders}
\begin{figure}[h]
\centering
\includegraphics[width=0.49\columnwidth]{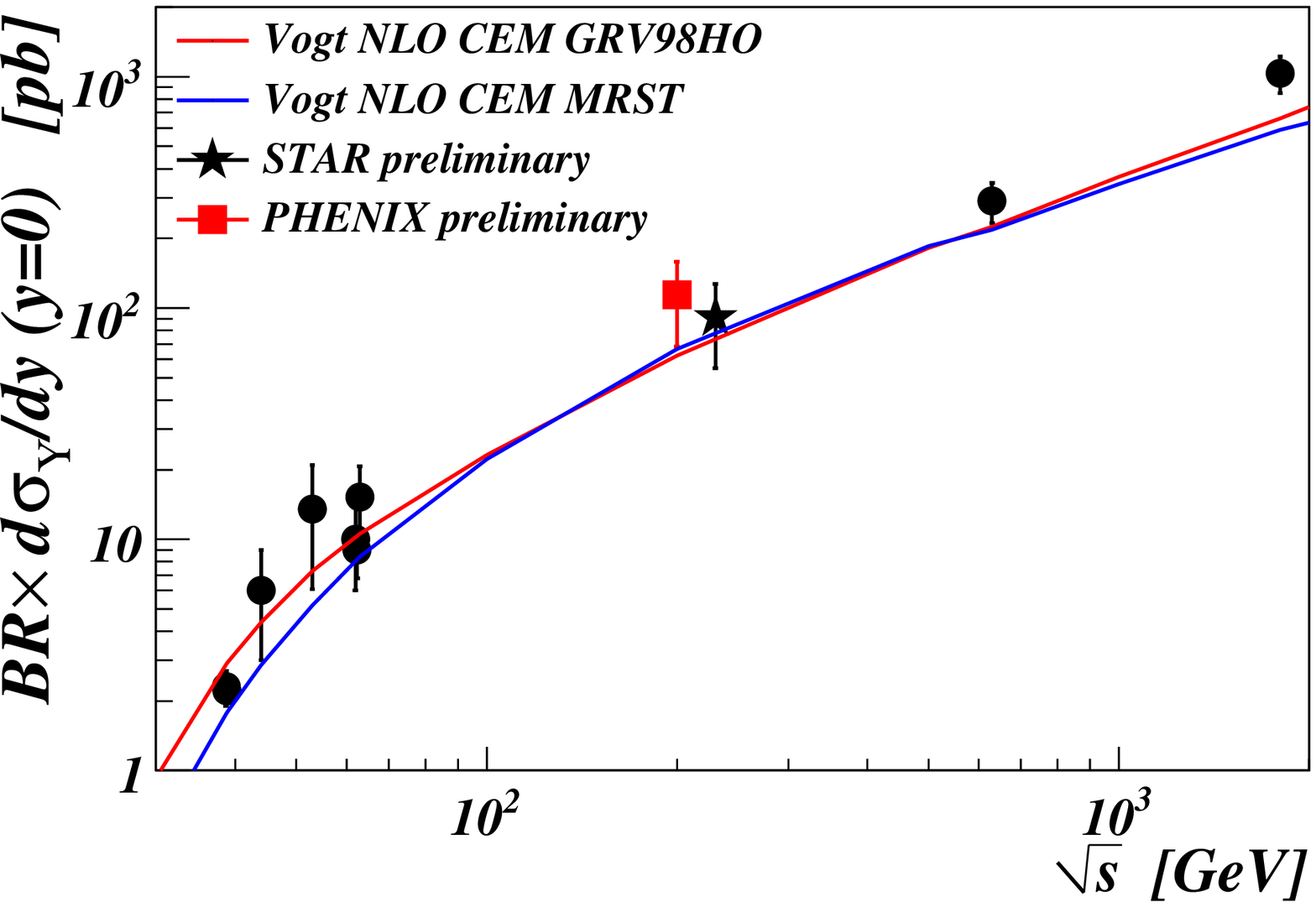}
\includegraphics[width=0.49\columnwidth]{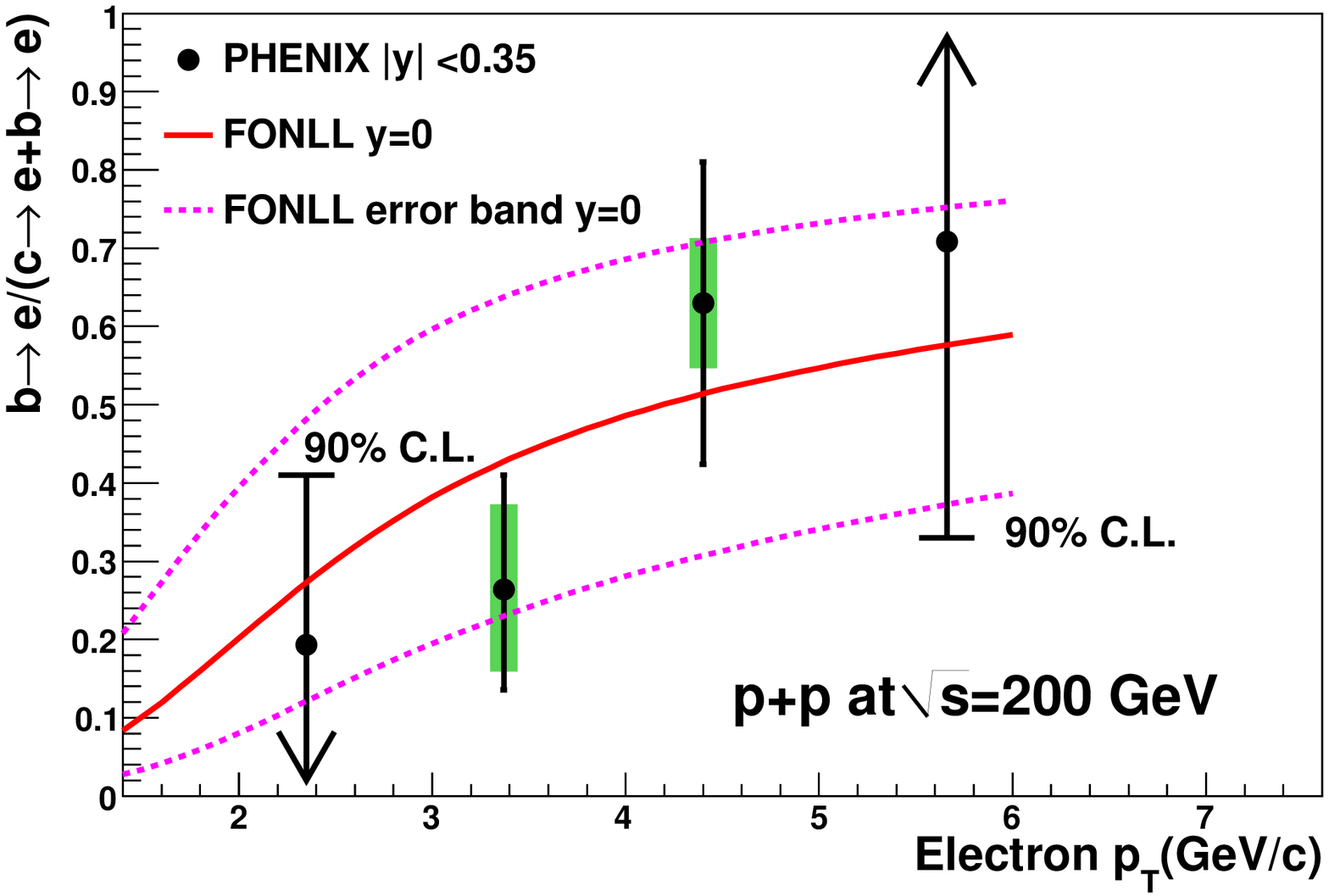}
\caption{Left -- cross section for $\Upsilon$ production times  
branching ratio as a function of centre-of-mass energy. PHENIX and  
STAR data points are shown as well as theory predictions. Right --  
bottom fraction as a function of electron $p_T$ measured in PHENIX  
data (points), compared to the FONLL prediction (solid line) and its  
uncertainty (dashed lines).}\label{Fig:Phenix}
\end{figure}

The era of ``beauty'' is opening at RHIC~\cite{RHIC}.
Heavy flavours are key tools to probe and understand independently of  
other methods the properties of the hot and dense nuclear medium  
through their energy loss. Both the STAR and PHENIX experiments  
reconstruct $\Upsilon \to e^+e^-$ signals in $\sqrt{s}=200$ GeV $pp$   
(Fig.~\ref{Fig:Phenix}, left) and $d+$Au collisions, and observe  
binary scaling of the production. Production of quarkonia is not  
modified in $d+$Au collisions (e.g. STAR measures $R_{d\mathrm{Au}} =  
0.98 \pm 0.32 (\mbox{stat}) \pm 0.28 (\mbox{syst})$) but appears  
suppressed in Au+Au collisions, which is not fully understood. In  
addition, both experiments find large $b$--quark contribution to the  
single-$e$ spectrum at $p_T>4$ GeV, in agreement with FONLL pQCD
  theory~\cite{Cacciari:2005rk} (Fig.~\ref{Fig:Phenix}, right).
  Next generation heavy ions experiments plan to improve these results  
and obtain a clearer picture: in just one month of running ALICE plans  
to collect large samples of \emph{onia}
with significances $S/\sqrt{S+B} \approx 10-100$. Similarly, CMS plans  
to efficiently reconstruct charmonia and bottomonia in Pb+Pb  
collisions and be able to determine the production cross section~ 
\cite{LHC}.

\section{News from HERA and $e^+e^-$ colliders}

\subsection{New results from $e^+e^-$ colliders}

An observation of an anomalous line-shape of the $e^+e^-\rightarrow\,$hadrons
cross sections near $3.770\,$GeV has been reported by BES~\cite{chen}.
It is inconsistent with only one $\psi(3770)$ state, suggesting
either a new structure in addition to the $\psi(3770)$ resonance or
some physics effects distorting the pure $D$-wave Breit-Wigner shape
of the cross sections. The observation suggests that
a surprisingly large non-$D{\bar D}$ branching fraction of the $\psi(3770)$
decays measured previously by BES~\cite{chen} may partially be
due to the assumption of only one simple resonance near $3.770\,$GeV.

Recent measurements of the leptonic and semileptonic $D^0$, $D^+$
and $D^+_s$ branching fractions have been reported by CLEO~\cite{chen}.
Using lattice QCD calculations, the $|V_{cs}|$ and $|V_{cd}|$
elements of the CKM matrix were determined and found to be in
agreement with previous measurements.

New BELLE measurement of the $X(3872)\rightarrow D^0{\bar D}^{*0}$
decay~\cite{kuhr}
has revealed the $X$ mass value of $3872.6^{+0.5}_{-0.4}\pm0.4\,$MeV
in fair agreement with the mass measurement in the dominant
$X(3872)\rightarrow J/\psi\pi\pi$ decay mode.
A first evidence for the $X(3872)\rightarrow \psi(2S)\gamma$ decay
reported by BABAR~\cite{kuhr} has allowed to determine the ratio
of the branching fractions
$${\cal B}_{X\rightarrow \psi(2S)\gamma}/{\cal B}_{X\rightarrow J\psi\gamma}=3.4\pm1.4\,,$$
which is above the molecule model expectation~\cite{swanson}.

The situation with the charged state $Z^+(4430)$ remains unclear~\cite{kuhr}.
BELLE has confirmed its observation
by the Dalitz plot analysis of the
$Z^+(4430)\rightarrow \psi(2S)\pi^+$ decay~\cite{kuhr}.
However, BABAR has found no evidence for the charged state
in the above decay and in the $J/\psi\pi^+$ final state.
Meanwhile, BELLE has reported two further charged states
$Z(4050/4250)^+\rightarrow\chi_{c1}\pi^+$.

BABAR has confirmed its observation
of the $\eta_b(1S)$ bottomonium ground state in the
decay $Y(3S)\rightarrow\gamma\eta_b(1S)$ by a new $\eta_b(1S)$
measurement in the $Y(2S)\rightarrow\gamma\eta_b(1S)$ decay~\cite{ziegler}.
Searches for the light Higgs-like particle in the $Y(2S)$ and $Y(3S)$
radiative decays have revealed no signal.

\subsection{New HERA results}

New measurement of $J/\psi$ production in proton-nucleus
collisions performed by HERA-B~\cite{spighi} has confirmed that
the $dN/dp_T$ distribution becomes  broader with increasing
atomic mass number, $A$. The $dN/dx_F$ distribution also tends
to become broader and its centre moves towards negative $x_F$
values with increasing $A$.
The fraction of $J/\psi$ mesons produced through the $\chi_c$ decay
has been measured to be
$R_{\chi_c}=18.8\pm1.3^{+2.4}_{-2.2}\,\%$
with a ratio of $\chi_{c1}$ and $\chi_{c2}$ contributions
$R_{\chi_{c1}}/R_{\chi_{c2}}=1.02\pm0.20$~\cite{spighi}.
The measured $J/\psi$ decay angular distributions
indicate the polar anisotropy with a preferred spin component 0
along the reference axis. The polar anisotropy increases with
decreasing $p_T(J/\psi)$~\cite{spighi}.
The $J/\psi$ helicity distributions measured by ZEUS in the inelastic
photoproduction regime~\cite{bertolin} have been compared to LO QCD predictions in
the colour-singlet, colour-singlet plus colour-octet and $k_T$
factorisation approaches; none of the predictions
can describe all aspects of the data.

\begin{wrapfigure}{r}{0.5\columnwidth}
\centerline{\includegraphics[width=0.49\columnwidth]{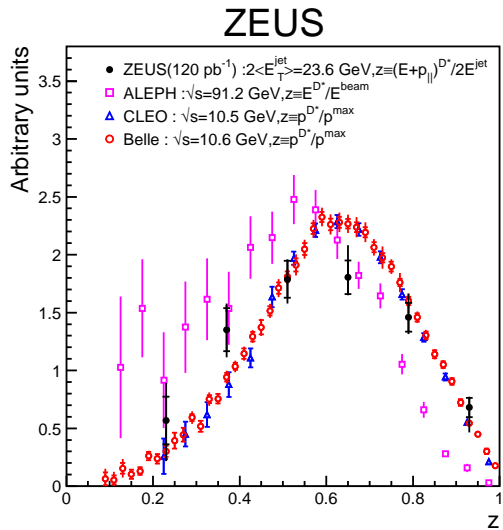}}
\caption{Charm fragmentation function in transition to $D^{*+}$
for the ZEUS data compared
to measurements of BELLE, CLEO and ALEPH.}\label{Fig:fragm}
\end{wrapfigure}
Charm fragmentation function in the transition from a charm quark
to a $D^{*+}$ meson measured
by ZEUS using the variable $z=(E+p_{||})^{D^{*+}}/2E^{jet}$~\cite{gladilin}
is compared in Fig.~\ref{Fig:fragm} with previous measurements
from BELLE, CLEO and ALEPH.
The corresponding scale of the ZEUS data is given by twice the average transverse energy of the jet, $23.6\,$GeV, and is between the two $e^+e^-$ centre-of-mass energies.
The ZEUS data in  Fig.~\ref{Fig:fragm} are shifted somewhat 
to lower values of $z$ compared to the CLEO and BELLE data with the ALEPH data
even lower, which is consistent with the expectations
from the scaling violations in QCD.
The value of the free parameter in the Peterson {\it et al.}
fragmentation function~\cite{Peterson:1982ak}
extracted from the ZEUS data within the framework  of the
next-to-leading order (NLO) QCD, $\epsilon=0.079\pm0.008^{+0.010}_{-0.005}$,
exceeds the value $0.035$ obtained from an NLO fit~\cite{nason}
to the ARGUS data.

Sizable production of the excited charm and charm-strange mesons has been
observed in $ep$ interactions by ZEUS~\cite{gladilin}.
The fractions of $c$ quarks hadronising into $D^0_1$, $D^{*0}_2$ or $D^+_{s1}$
mesons are consistent with those obtained in $e^+e^-$ annihilations.
The measured $D^0_1$ width,
$\Gamma(D^0_1)=53.2\pm7.2^{+3.3}_{-4.9}\,$MeV, is above
the world average value. This is possibly 
due to a larger admixture from the broad $S$-wave decay
at ZEUS compared to the measurements with
restricted phase space~\cite{gladilin}.

\begin{wrapfigure}{r}{0.5\columnwidth}
\centerline{\includegraphics[width=0.49\columnwidth]{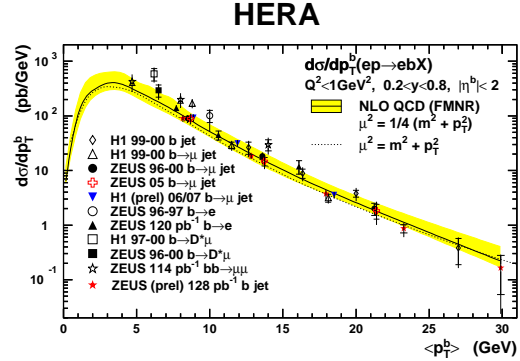}}
\caption{Summary of HERA differential cross sections for $b$-quark production
in $ep$ interactions in photoproduction regime as function of
$p^b_T$ as measured by H1 and ZEUS.}\label{Fig:herabb}
\end{wrapfigure}

New measurements of beauty dijet photoproduction
have been performed by H1~\cite{list},
using events with a muon in the final state,
and ZEUS~\cite{miglioranzi}.
The production cross sections were found to be compatible with the previous
HERA measurements and with NLO QCD predictions.
The measured cross sections translated
into a differential cross section $\frac{d\sigma}{dp^b_T}$
in the pseudorapidity range $|\eta_b|<2$
are compared to the
previous HERA measurements and
the FMNR NLO QCD
predictions~\cite{fmnr} in Fig.~\ref{Fig:herabb}.

Inclusive production of $D^{*\pm}$ mesons in deep inelastic scattering (DIS)
has been measured by H1 in two ranges of the exchanged photon virtuality,
$5<Q^2<100\,$GeV$^2$~\cite{jung} and $100<Q^2<1000\,$GeV$^2$~\cite{brinkmann},
using 2004-2007 data corresponding to an integrated luminosity
of $350\,$pb$^{-1}$. The data are described reasonably well
by the NLO calculation HVQDIS~\cite{hvqdis}.
The data description by the leading-order Monte-Carlo simulations RAPGAP
and CASCADE~\cite{cascade} is satisfactory only in the low-$Q^2$ range.
The measured $D^{*\pm}$ cross sections were used to extract the charm
contribution to the proton structure, $F_2^{c{\bar c}}(x,Q^2)$,
where $x$ is the Bjorken scaling variable.
The $F_2^{c{\bar c}}(x,Q^2)$ values in the frameworks of
DGLAP and CCFM evolutions were obtained using
for extrapolation HVQDIS and CASCADE, respectively.

H1 has also measured the inclusive charm cross sections simultaneously
with the inclusive beauty cross sections in the range $5<Q^2<650\,$GeV$^2$
using the impact parameters and the secondary vertex position
reconstructed with the H1 vertex detector~\cite{thompson}.
To obtain fractions of charm, beauty and light quarks in
the inclusive sample a neural network was used.
The measured cross sections are described reasonably well
by the predictions based on the DGLAP and CCFM evolutions.
The charm and beauty contributions to the proton structure,
$F_2^{c{\bar c}}(x,Q^2)$ and $F_2^{b{\bar b}}(x,Q^2)$,
were extracted in the framework of DGLAP evolution.
To gain in precision,
the extracted $F_2^{c{\bar c}}(x,Q^2)$ values were combined with those
obtained using the cross sections of $D^{*\pm}$ meson
production~\cite{thompson}.
The measurements were interpolated to the common $x$, $Q^2$ grid and
averaged using the procedure developed for the inclusive $F_2$
taking into account correlations.

\begin{figure}[!ht]
\begin{center}
\centerline{\epsfxsize=2.7in\epsfbox{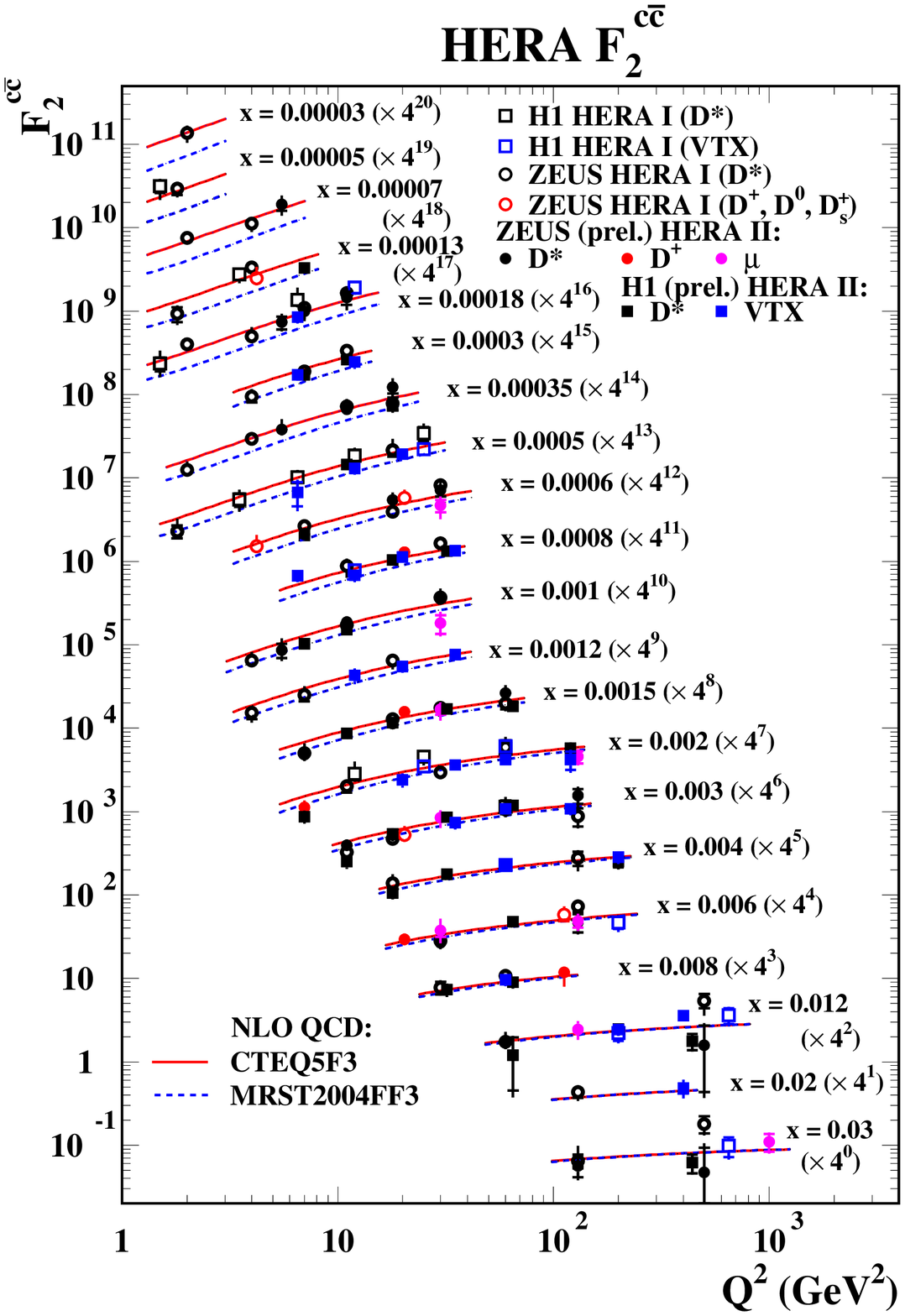}
\epsfxsize=2.7in\epsfbox{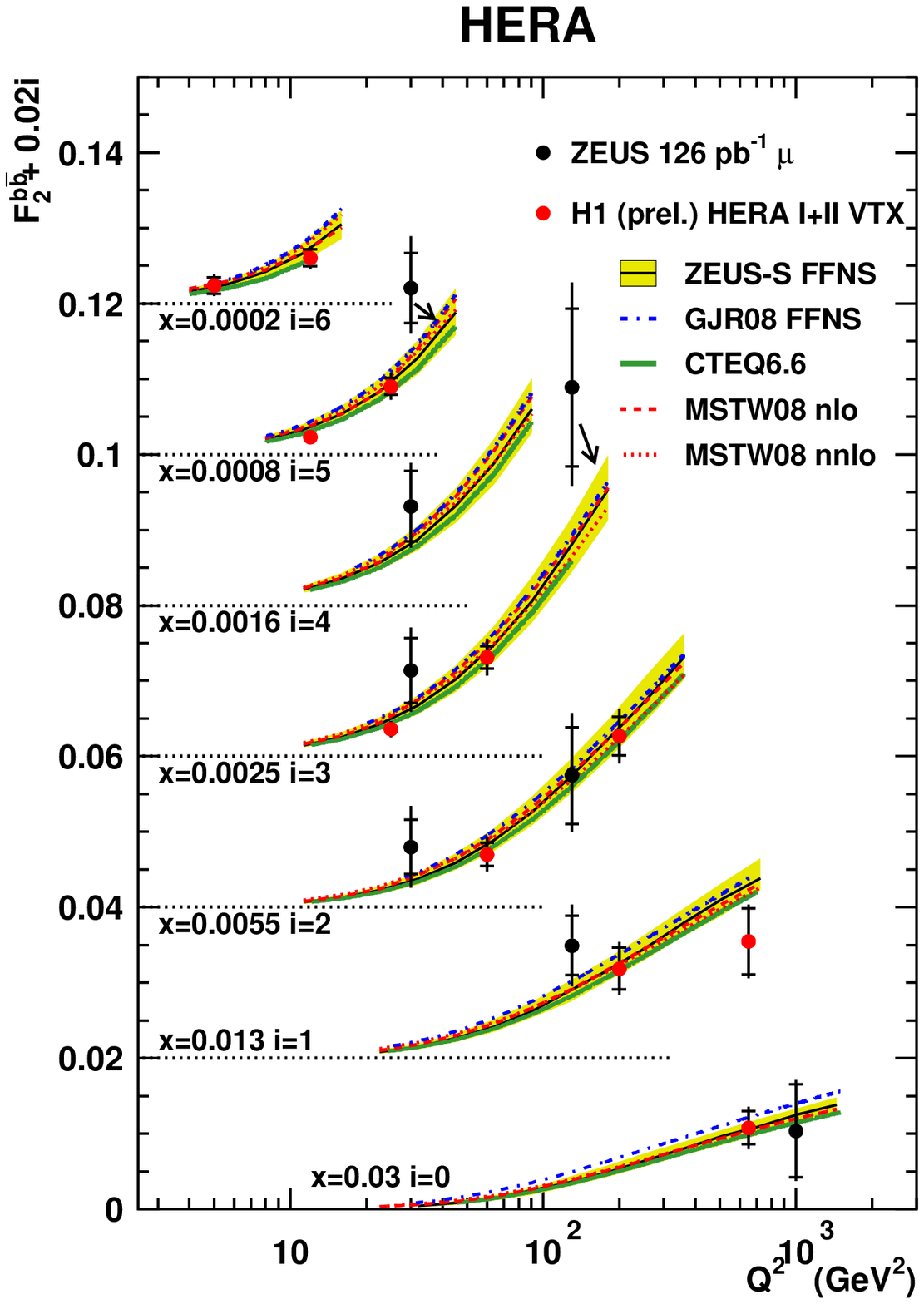}}
\end{center}
\caption[*]{Summary of HERA measurements of the structure functions
(left) $F_2^{c{\bar c}}$ and (right) $F_2^{b{\bar b}}$
plotted as functions of $Q^2$ for fixed $x$ values.}\label{Fig:f2ccbb}
\end{figure}

The production of $D^\pm$ and $D^0$ mesons in DIS has been measured
by ZEUS in the range $5<Q^2<1000\,$GeV$^2$
using the ZEUS microvertex detector to reconstruct displaced
secondary vertexes~\cite{roloff}.
The measured cross sections were found to be
in agreement with the predictions of NLO QCD
with the proton parton density functions extracted from
inclusive DIS data.
The measured $D^{\pm}$ and $D^  0$ cross sections were used
to extract $F_2^{c{\bar c}}(x,Q^2)$ within the framework of DGLAP
NLO QCD. The extracted $F_2^{c{\bar c}}(x,Q^2)$ values agree well
with the previous measurements obtained using
cross sections of $D^{*\pm}$ meson production.

ZEUS has also measured the production of charm and beauty quarks
in DIS for $Q^2>20\,$GeV$^2$ using the heavy-quark decays into
muons~\cite{roloff,bindi}.
The fractions of muons originating from charm and beauty
decays were determined using the muon momentum
component transverse to the axis of the associated jet,
the distance of closest approach of the muon track to the centre
of the interaction region in the transverse plane
and the missing transverse momentum parallel to the muon direction.
The latter requirement was important for distinguishing
contributions from charm and light flavours.
The measured muon differential cross sections
were compared to the NLO QCD calculations;
the agreement was found to be good for charm 
and reasonable for beauty.
The $F_2^{c{\bar c}}(x,Q^2)$ and $F_2^{b{\bar b}}(x,Q^2)$
values were also measured and found to agree well with other
measurements based on independent techniques.
The HERA measurements of $F_2^{c{\bar c}}$ and $F_2^{b{\bar b}}$
are summarised in Fig.~\ref{Fig:f2ccbb}.

\section*{Acknowledgements}

We thank all speakers of the working group ``Heavy flavours''
for their talks and participation in discussions.
Special thanks are due to the DIS\,2009 organisers for their
hospitality and help in running the sessions smoothly.


\begin{footnotesize}



%

\end{footnotesize}


\end{document}